\documentclass{article}

\usepackage[final]{neurips_2019}

\usepackage[utf8]{inputenc} 
\usepackage[T1]{fontenc}    
\usepackage{hyperref}       
\usepackage{url}            
\usepackage{booktabs}       
\usepackage{amsfonts}       
\usepackage{nicefrac}       
\usepackage{microtype}      
\usepackage{enumitem}
\usepackage{authblk}
\usepackage{amsmath}
\usepackage{caption}
\usepackage{subcaption}
\usepackage{booktabs}
\usepackage{multirow}
\usepackage{amssymb}
\usepackage{booktabs}
\usepackage{graphicx}

\title{BERT Learns (and Teaches) Chemistry\vspace{-0em}}

\author[1,2]{Josh Payne}
\author[1,2]{Mario Srouji}
\author[1,2]{Dian Ang Yap}
\author[1,2]{Vineet Kosaraju}
\affil[1]{Department of Computer Science, Stanford University}
\affil[2]{\{joshp007, msrouji, dianang7, vineetk\}@stanford.edu}

\begin{document}

\maketitle


\vspace{-1.5em}
\begin{abstract}
Modern computational organic chemistry is becoming increasingly data-driven. There remain a large number of important unsolved problems in this area such as product prediction given reactants, drug discovery, and metric-optimized molecule synthesis, but efforts to solve these problems using machine learning have also increased in recent years. In this work, we propose the use of attention to study functional groups and other property-impacting molecular substructures from a data-driven perspective, using an transformer-based model (BERT) on datasets of string representations of molecules and analyzing the behavior of its attention heads. We then apply the representations of functional groups and atoms learned by the model to tackle problems of toxicity, solubility, drug-likeness, and synthesis accessibility on smaller datasets using the learned representations as features for graph convolution and attention models on the graph structure of molecules, as well as fine-tuning of BERT. Finally, we propose the use of attention visualization as a helpful tool for chemistry practitioners and students to quickly identify important substructures in various chemical properties.
\end{abstract}

\section{Introduction}
There are several problems in organic chemistry, such as the prediction of chemical properties of unknown molecules, that are very difficult and remain largely unsolved. Handcrafting approaches are not scalable as the number of reported organic reactions constantly increases and significant time investment is needed to keep up with the literature. However, certain heuristics to discover new molecules have proven useful for these tasks, such as the notion of a functional group. Functional groups are sets of connected atoms that determine properties and reactivity of a parent molecule. They are a cornerstone of modern organic chemistry and chemical nomenclature. There are many scientific papers that focus on properties and reactivity of various functional groups, however there is little attention paid to the study of functional groups from a data-driven point of view.

In modern applications such as drug discovery, the search space is estimated to be in the range of $10^{23}$ to $10^{60}$ molecules \cite{polishchuk2013estimation}; however, functional groups and property-impacting substructures can greatly assist in restricting the size of this space, as they indicate the presence of specific chemical properties in molecules. Understanding the influence of these functional groups in relation with each other is also a crucial step in understanding molecular properties such as stability and reactivity, besides providing insights into activating/deactivating properties when we seek to synthesize new molecules. They remain a widely researched topic in chemistry, ranging from understanding activating/deactivating groups, steric hindrance, and resonance in large molecules.

Currently, the majority of theoretical studies utilize functional groups as a rule-based definition to categorize compounds into classes, allowing chemists to perform large-scale automated chemical classification based on a structure-based chemical taxonomy. Additionally, while machine learning has been applied to predict reactions, and to generate new molecular graphs, little work has been done to either leverage functional groups to perform predictions or to interpret the deep learning models to identify functional groups (hence allowing the groups to emerge from the data as opposed to from heuristics).

In this work we propose a method to perform automated identification of important molecular substructures without any chemistry domain expertise, hard-coding of structures, heuristic methods, pre-selection approaches, or restricted definitions. We instead detect important functional groups of atoms in molecules by allowing them to emerge from the data. In this way, we aim to learn these rules of chemistry directly. We do this through the use of attention mechanisms on the SMILES strings of molecules, by setting up the task of identifying important substructures in BERT's pretraining procedure and a following task of fine-tuning BERT for regression on solubility, drug-likeness, and synthesis accessibility, in order to identify important substructures of molecules. 

\section{Related Work}
\textbf{Language Modeling} \\
BERT (\textbf{B}idirectional \textbf{E}ncoder \textbf{R}epresentations from \textbf{T}ransformers) \cite{bert} was demonstrated to outperform previous language models on a variety of tasks with high efficiency. Other works also explore multi-task learning on language models that leverage large amounts of cross-task data, which also exhibit from a regularization effect that leads to more general representations to help adapt to new tasks and domains \cite{liu2019multi}. Moreover, recent work and tools have shown that interpretability is tractable with BERT \cite{Clark_2019, Hoover2019exBERTAV}.
 
BERT builds upon the encoder structure of the encoder-decoder architecture of transformer, which uses multi-head self attention \cite{bahdanau2014neural, bau2018identifying}. Different techniques have been developed to better interpret attention maps, such as attention matrix heatmaps \cite{bahdanau2014neural, rush2015neural, rocktaschel2015reasoning} and bipartite graph representations \cite{liu2018visual, strobelt2018s}. 
 
\textbf{Multitask and Meta-Learning} \\
Although the field of meta learning is relatively nascent, there are still several technical approaches that have been utilized on similar problems that we believe can be applied to computational chemistry. For instance, Ramsundar et. al. \cite{massively_multitask} were able to demonstrate how increasing the amount of data and tasks can greatly increase performance when discovering novel drugs; this parallels our initial hypothesis with discovering functional groups. Further, Ren et. al. \cite{ren18fewshotssl} augmented prototypical networks for semi-supervised classification; we believe such an approach may also help us with classifying and grouping functional groups and molecules. Finally, Garcia et. al. \cite{garcia2017few} applied few shot learning to GNNs; as molecules and reactions can be modeled using graphs and hypergraphs, we believe a similar approach may be useful here.

\textbf{Using Machine Learning for Chemical Inference Tasks} \\
Previous work leveraged a transformer model to perform chemical reaction prediction \cite{schwaller2018molecular}. Given reactants and reagents, they predict the products, and similar to other work, they treat reaction prediction as a machine translation problem between the SMILES strings. They show that a multi-head attention Molecular Transformer model outperforms all algorithms in the literature, requiring no handcrafted rules. They did not however leverage functional groups in their predictions, or attempt to interpret the model to detect groupings of atoms, and their effects. 

Another work by You et al. \cite{you2018graph} proposes the Graph Convolutional Policy Network (GCPN), an approach to generate molecules where the generation process can be guided towards specified desired objectives, while restricting the output space based on underlying chemical rules. Graph representation learning is used to obtain vector representations of the state of generated graphs, adversarial loss is used as reward to incorporate prior knowledge specified by a dataset of example molecules, and the entire model is trained end-to-end in a reinforcement learning framework.

One slightly related work creates molecule Deep Q-Networks (MolDQN), for molecule optimization by combining domain knowledge of chemistry, and reinforcement learning techniques \cite{zhou2019optimization}. They define modifications on molecules to ensure chemical validity. MolDQN achieves comparable or better performance against several other recently published algorithms for benchmark molecular optimization tasks. However, they argue that many of these tasks are not representative of real optimization problems in drug discovery. They extend their model with multi-objective reinforcement learning, to maximize drug-likeness while maintaining similarity to the original molecule. In both of the above works, we hypothesize that the attention head behavior can be used to make the reward function more dense, decreasing training time.

Regression tasks on chemical attributes such as toxicity, solubility (logP), drug-likeness (QED), and synthesis accessibility (SAS) have been proposed by several works. Graph Neural Networks (namely, Graph Convolutional Networks and Graph Attention Networks) have been shown to be effective on toxicity tasks such as the Tox21 dataset \cite{gcn, gat}, and recently, pretraining procedures for graph neural networks have been shown to be effective on chemical inference \cite{Hu2019PretrainingGN}. Explainability and identification of important substructures on molecule graphs have also been recently proposed in \cite{xiong2019pushing} using GATs. However, these graph-based machine learning approaches are inherently limited in their depth, as too many layers will cause oversmoothing. This issue doesn't exist with deep language models such as BERT, which is built using multiple transformers and hundreds of millions of parameters.

However, there exists the issue of learning the encoding rules built into SMILES strings, which are formally defined in Section \ref{approach}. One problem that arises here is that two unique SMILES strings representing the same molecule can have different embeddings. How can a language model be trained to recognize isomorphism between unique, yet equivalently mapping, SMILES strings, essentially learning a truer representation of the molecule the SMILES string encodes? One work proposes a very elaborate and hand-designed architecture that performs well on regression tasks using SMILES string \cite{alperstein2019smiles}. However, this architecture, while achieving state-of-the-art results, is not a clean solution that allows the encoding rules to arise from the data or training procedure.

\textbf{Detecting Functional Groups within Molecules} \\
There currently exists a small amount of work on identifying functional groups via an automated process, as in \cite{ertl2017algorithm}. The approach of this paper is based on processing hetero-atoms and their environment with the addition of some other functionalities, like multiple carbon–carbon bonds. However while this is a step forward from manual curation, the algorithm involves an iterative process of marking and merging of atoms based on pre-defined heuristics. This inherently restricts the ability to learn functional groups through data, or through chemical reaction properties. Additionally, the work admits that this algorithm does not provide an ultimate definition of functional groups, as every chemist has a slightly different understanding about what a functional group is. For example, in their present algorithm they restricted their definition only to classical acetal, thioacetal or aminal centers, and did not consider other similar systems, like alpha-substituted carbonyls, or similar bonds. Hence the method they used also suffers from a strict formulation of what consititutes a functional group within a molecule. 

Another state-of-the-art program for identification of functional groups is called checkmol \cite{haider2003checkmol}. Checkmol is a command-line utility program which reads molecular structure files in different formats, and analyzes the input molecule for the presence of various functional groups and structural elements, from a predefined set of 204 functionalities (through pattern matching). The philosophy of checkmol is slightly different from that of their approach, as checkmol defines functional groups in a hierarchical manner. In summary, both approaches had very similar results for the functional groups detected. This shows that systems based on a well selected list of substructures (as apparently checkmol is) can provide useful information about the functional group composition of general molecular data-sets, however both approaches involved expert knowledge in the field of chemistry in order to either manually curate a selection of functional groups, or to create a strictly defined heuristic to detect functional groups iteratively. 

There have been a few machine learning-based approaches in identifying functional groups from data. Last month, Pope et al. \cite{fggnn} released a paper outlining a GCN-based approach to detecting functional groups in data. This work was an extension of their 2019 paper on explainability methods for GCNs \cite{pope2019explainability} and proposed the application of contrastive gradient-based saliency maps, class activation mapping (CAM), and excitation backpropagation, previously proposed for convolutional neural networks, to the graph convolutional neural network domain. However, GCNs are foundationally less expressive than many modern sequence models, as their depth is restricted by the number of layers they can have before oversmoothing occurs. Additionally, they are not as readily explainable as sequence models with attention layers.

\section{Approach}\label{approach}
\subsection{Datasets}
Some molecules are more readily synthesizable than others; that is, the chemical properties present in synthesizable molecules may provide a stronger indication of functional groups. The first dataset we analyze is a set of 250,000 readily synthesizable molecules collected from the ZINC15 database\footnote{Link to the ZINC15 250k dataset: \url{https://github.com/aspuru-guzik-group/chemical_vae/tree/master/models/zinc}}. This dataset contains the SMILES string for each of these molecules, as well as the logP, QED, and SAS values. 
The term SMILES (Simplified Molecular Input Line Entry System) refers to a string notation for encoding molecular structures \cite{toropov2005simplified}. 
In order to more easily perform computation and analysis on molecules, David Weininger developed the SMILES (Simplified Molecular Input Line Entry System) specification at the USEPA Mid-Continent Ecology Division Laboratory in Duluth in the 1980s. 

Typically, numerous equally valid SMILES strings can be written for a molecule. SMILES contains the same information as might be found in an extended connection table. The primary reason SMILES is more useful than a connection table is that it is a linguistic construct, rather than a computer data structure. SMILES is a true language, with a simple vocabulary (atom and bond symbols) and only a few grammar rules. SMILES notation consists of a series of characters containing no spaces. Hydrogen atoms may be omitted (hydrogen-suppressed graphs) or included (hydrogen-complete graphs). 

In terms of a graph-based computational procedure, SMILES is a string obtained by printing the symbol nodes encountered in a depth-first tree traversal of a chemical graph. The chemical graph is first trimmed to remove hydrogen atoms and cycles are broken to turn it into a spanning tree. Where cycles have been broken, numeric suffix labels are included to indicate the connected nodes. Parentheses are used to indicate points of branching on the tree. Below is an example SMILES string and corresponding molecule:

\begin{center}
\texttt{CC(C)(C)c1ccc2occ(CC(=O)Nc3ccccc3F)c2c1}\\
\begin{center}
\includegraphics[width=150px]{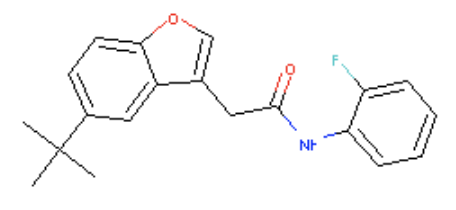}
\end{center}
\end{center}

After studying chemical properties with attention in the ZINC15 250k dataset, we want to apply our new representations to a smaller, more difficult task. The Tox21 dataset is a multitask dataset of 12 unique toxicity tasks, each a binary categorical task. It contains roughly 7,000 molecules and typical featurizations include atomic number, aromicity, donor status, and acceptor status.

\subsection{Neural Machine Translation and Language Modeling}

Neural machine translation was proposed by Kalchbrenner and Blunsom (2013), Sutskever et al. (2014) and Cho et al. (2014b). An encoder neural network reads and encodes a source sentence into a fixed-length vector. A decoder then outputs a translation from the encoded vector. The whole encoder–decoder system is jointly trained to maximize the probability of a correct translation given a source sentence. 

Recurrent neural networks (RNN) \cite{mikolov2010recurrent}, and long short-term memory (LSTM) \cite{hochreiter1997long} have been established as the state of the art approaches in sequence modeling, and are commonly used in Neural Machine translation. Recurrent models factor computation along the symbol positions of the input and output sequences, by aligning the positions to steps in computation time. They generate a sequence of hidden states $h_t$, as a function of the previous hidden state $h_{t-1}$, and the input for position $t$. The issue is that a neural network needs to be able to compress all the necessary information of a source sentence into a fixed-length vector. This may make it difficult for the neural network to cope with long sentences, especially those that are longer than the sentences in the training corpus. Cho et al. (2014b) showed that the performance of a basic encoder–decoder deteriorates rapidly as the length of an input sentence increases. 

In order to address this issue, attention mechanisms \cite{bahdanau2014neural} were introduced as an extension to the encoder–decoder model. The generic attention takes in three values: a value (V), a key (K), and a query (Q). From these values, it applies a standard dot-product attention, as shown in Equation \ref{eqn1}, with a scaling factor of $\frac{1}{\sqrt{d_k}}$ added to ensure that the value of the dot product doesn't grow too large in magnitude with respect to $d_k$, the dimension of the key.
\begin{equation}
\label{eqn1}
    \textrm{Attention}(Q, K, V) = \textrm{softmax}\left(\frac{QK^T}{\sqrt{d_k}}\right) V
\end{equation}

Attention models also have high interpretability, allowing the researcher to observe what the attention heads are "attending" to by investigating the values of the attention heads as inputs are fed through, as done in BERT.

\subsection{Multi-Task Learning}
In this work, we propose a tripartite approach to studying and understanding functional groups and applying these findings to studying chemical reactions and properties. These datasets will be investigated after we've learned to learn representations of functional groups from the larger datasets.

To create meaningful representations of atomic tokens in the context of molecules and reaction networks, we intend to train BERT on several large reaction datasets, including the ZINC15 250k dataset. A multi-task learning approach will be used: be learning ``masked" molecules in string representations, relations learned between molecular string representations will hopefully pave the way for downstream tasks such as link prediction, functional group identification and generative tasks. Moreover, we hope to achieve SoA accuracy on reaction prediction tasks with these datasets.

We want to explore the explainability and interpretability of this model and use it to identify functional groups. Our goal is to be able to use these predictions to validate the model's interpretability using existing knowledge about functional groups as well as identify new, useful functional groups for chemists. Our hypothesis is that an analysis of the behaviour of BERT's attention heads will give indications of certain atoms which form functional groups. Interpretability of multi-task learning and meta-learning remains an intriguing field that would pave better understanding with more research.

\begin{figure}[!htb]
    \centering
    \includegraphics[width=0.5\linewidth]{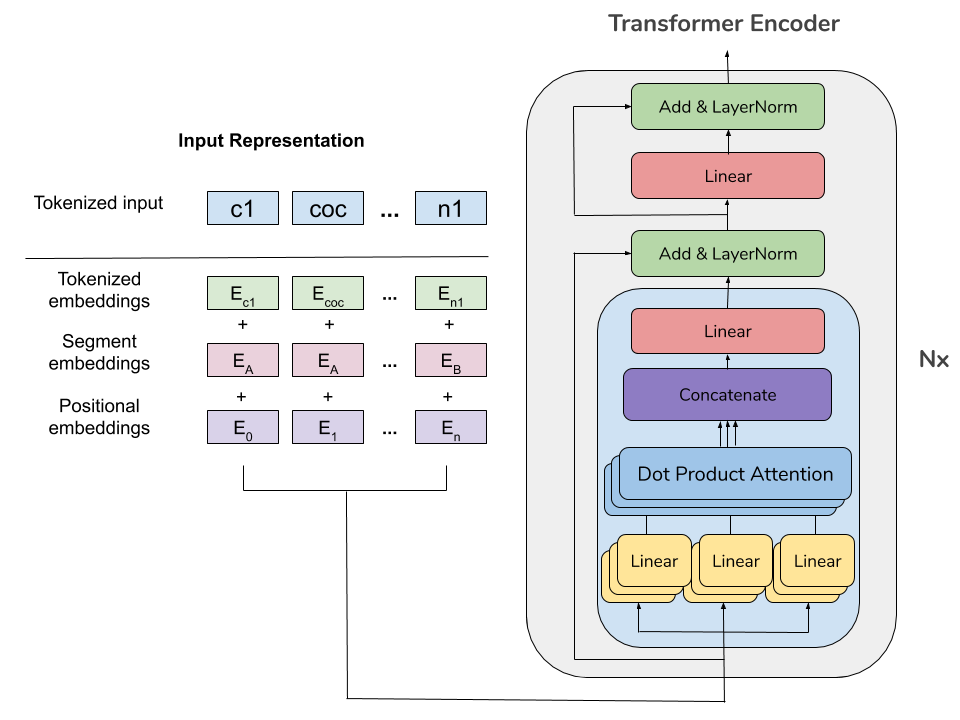}
    \caption{Visualization of BERT trained on SMILES strings. SMILES sequences are tokenized into functional groups, and the encoded embeddings with positional and token embeddings are fed in as input to BERT, the encoder structure of transformer. The light blue box denote attention mechanisms used in BERT.}
    \label{fig:fig1_transformer}
\end{figure}

Finally, we would like to transfer this knowledge to tasks on smaller, more difficult datasets, such as the Tox21 \cite{tox21} and MUV \cite{muv} multitask datasets. We intend to compare results of using learned functional group knowledge explicitly, features computed by BERT, combined with various language and graph learning models (LSTM \cite{hochreiter1997long}, BERT \cite{bert}, Transformer \cite{transformer}, GCN \cite{gcn}, GraphSAGE \cite{graphsage}, and Graph Attention Networks \cite{gat}, as molecules may be represented as SMILES strings and graphs equivalently). We'd also like to explore meta-learning techniques such as MAML \cite{maml} for improved performance in low-data settings.
 
\section{Experimental Setup}

\subsection{Pre-analysis using \textit{n}-grams}

Before training our BERT model, we performed a pre-analysis of the ZINC15 250k dataset using $n$-gram (or length $n$ subsequence) frequency evaluation in an attempt to find the most common atomic motifs within the molecules. Our hypothesis was that common motifs in synthesizable molecules correspond often to functional groups, as their synthesizability lends them nice reactive properties. An $n$-gram, in this context, is defined as a continuous sequence of atoms and bonds in the SMILES string representation of the molecule. For each molecule, we gathered all possible contiguous sub-sequences (of all possible lengths based on the SMILES string), by varying our $n$-gram "window" size. We then created a histogram of how frequently each SMILES sub-sequence (gathered through $n$-grams) appears across all of the molecules in the dataset. What we found is that common functional groups indeed appear with high frequency in this dataset; for instance, the 6-atom benzine ring appears in the top 50 most frequent $n$-grams.

\subsection{BERT learns Chemistry}

From the raw datasets, we tokenize using the regular expression proposed by \cite{molecular_transformer}. We also add tokens such as \texttt{[UNK], [SEP], [MASK], [PAD]} to denote special tokens such as unknown tokens, separation, mask and padding tokens. Similar to BERT trained on natural language models that mask out 15\% of the tokens, we mask out 15\% of the tokens in each SMILES string with a maximum sequence length of 128, which the model then learns to predict the masked out atoms and functional groups. This is equivalent to the pretraining step where by learning to predict the masked token, the molecular BERT could learn transferable tasks such as property prediction.

We employ a bidirectional context to learn contextual representations on molecular corpus. We also encode positional embeddings of length 512 on the tokenized inputs to learn relative positioning, and employ layer normalization and residuals to propagate signals through the deep 12-layer network. feed-forward sublayers within each layer computes non-linear hierarchical features, with 10\% dropout probability to improve generalization. We use 12 attention heads for the self-attention layers, which are then extracted for interpretable attention visualization.

\textbf{Training.} We train the model with a batch size of 128 on a learning rate of $2\times10^{-5}$ for 500,000 steps, with a linear warmup followed by Adam optimizer, where the initial linear warmup addresses the large variance in early learning stages \cite{liu2019variance}. This was conducted on a Google TPU over the course of 24 hours. At this point, the loss had converged to nearly zero.

\subsection{Analysis of BERT's Attention}
 Leveraging the self-attention mechanism, we visualize attention structures of BERT trained on chemical molecular representations such as SMILES, which could uncover new relationships on how different functional groups within a molecule attend and influence each other.
 

The BERT attention layer takes in an input of vector sequence $h = [h_1, h_2, \dots, h_n]$ corresponding to the $n$ tokenized functional groups of the SMILES sequence. Each vector $h_i$ is transformed into query, key and value vectors ($q_i, k_i, v_i)$ through separate affine transformations. The head computes attention weights between all pairs of tokens as softmax-normalized dot products between the query and key vectors. For token $i$ and for value vector $j$,

\begin{equation}
    \alpha_{ij} = \frac{\exp{q_i^T k_j}}{\sum_{n=1}^n \exp(q_i^T k_l)}
\end{equation}

The output $o_i$ of the attention head $i$ is a weighted sum of the value vectors, which represents how important every other tokenized functional group is when producing the next representation for the current tokenized functional group.
\begin{equation}
    o_i = \sum_{j=1}^n \alpha_{ij} v_j
\end{equation}

While previous work on visualizing attention were mostly on encoder-decoder models, we tweaked the classic transformer visualization to visualize attention maps on the encoder-only BERT. For a given transformer layer, we could visualize how a molecule, in its SMILES representation, attends to itself or to another molecule. Moreover, since parameters are not shared across the attention heads, each attention head is capable of producing a unique attention pattern. The attention heads are capable of capturing relative positional patterns and specific structural (lexical) patterns across atoms in molecules.


\begin{figure}
\centering
\begin{subfigure}{.45\textwidth}
  \centering
  \includegraphics[width=.7\linewidth]{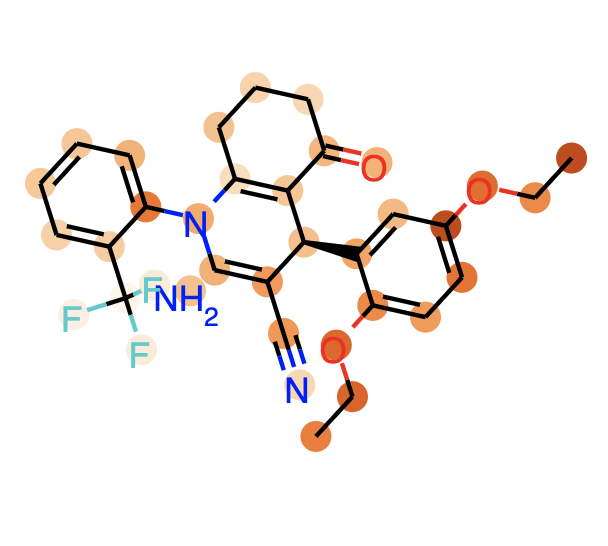}
  \caption{LogP, or solubility, is highly impacted by the length of carbon chains. We noted that several attention heads of BERT fine-tuned on logP value in the 10th layer attend heavily to carbon chains in this molecule.}
  \label{fig:sub1}
\end{subfigure}%
\hspace{.05\textwidth}
\begin{subfigure}{.45\textwidth}
  \centering
  \includegraphics[width=.7\linewidth]{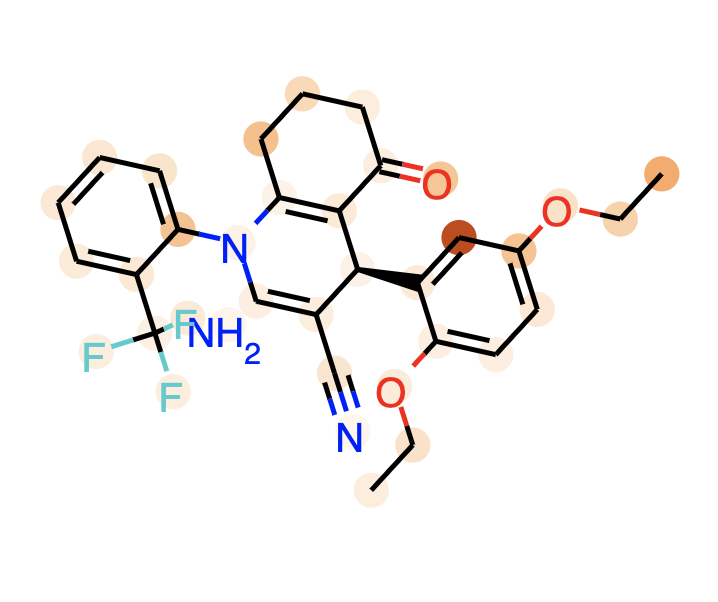}
  \caption{The attention heads of BERT pretraining attention does not pay special attention to carbon chains or other particular substructure.}
  \label{fig:sub2}
\end{subfigure}
\caption{The attention heads focus more heavily on the carbon chains; a carbon chain's length can help determine solubility.} 
\label{fig:test1}
\end{figure}

\section{Attention-based Analysis of Chemically-Attributed Molecular Substructures}
For a molecule $m$, we define atom $a_m$ as atom $a$ in molecule $m$. We would like to investigate which atom $a_m$ in molecule $m$ would react with other molecules $m'$, where $m' \in \{M\setminus m\}$, i.e. to identify the most likely site of $m$ where other molecules will react with. This part is done after pretraining the BERT structure on the chemical molecule datasets in an unsupervised manner by filling in the masked tokens as illustrated in the pretraining step above. Define the function $f_\theta : (M_1, M_2) \longrightarrow  \mathbb{R}^n$ denote the attention from  molecule $m \in M_1$ to $m' \in M_2$ parameterized by $\theta$, which are BERT's parameters. We have the extracted attention $a_m$ as

\begin{align}
    \tilde{a_m} &= \sum_{m' \in \{M\setminus m\}}\; \sum_{a_i \in m'}\; f_\theta(a_i, a_m) \\
    a_m &= \frac{\tilde{a}_m - \min(a_m)}{\max(a_m) - \min(a_m)}
\end{align}

Surprisingly, despite only being trained on SMILES string representation and learning the structure of sequence of tokenized atoms, without any data on reactions, BERT's attention heads learns to put more weight to sites which are, in reality, active functional groups that act as a site for reaction. Consider the example in Figure \ref{fig:reaction}:
\begin{figure}[!htb]
    \centering
    \includegraphics[width=\linewidth]{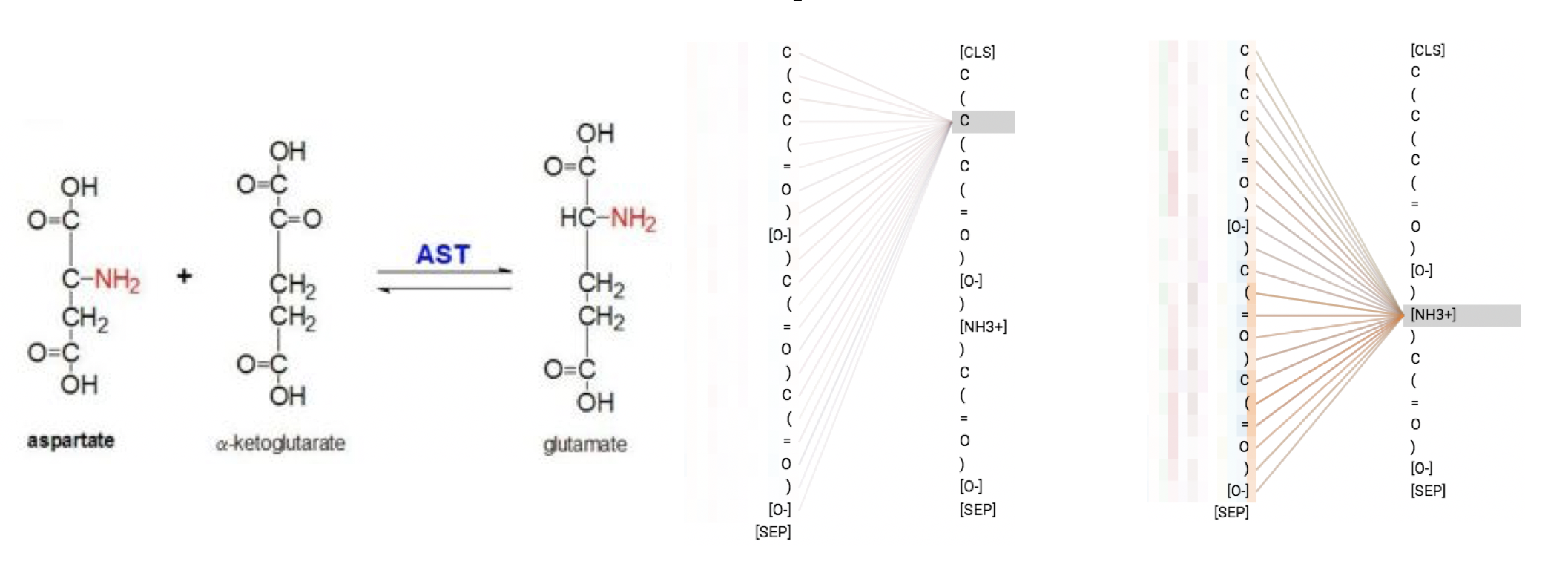}
    \caption{Left: illustration of chemical reaction on how aspartate reacts with $\alpha$-ketoglurate to form glutamate. Right: 
    Visualization of layer 3 of attention heads in BERT on different molecules in aspartate.}
    \label{fig:reaction}
\end{figure}

\begin{figure}[!htb]
    \centering
    \includegraphics[width=\linewidth]{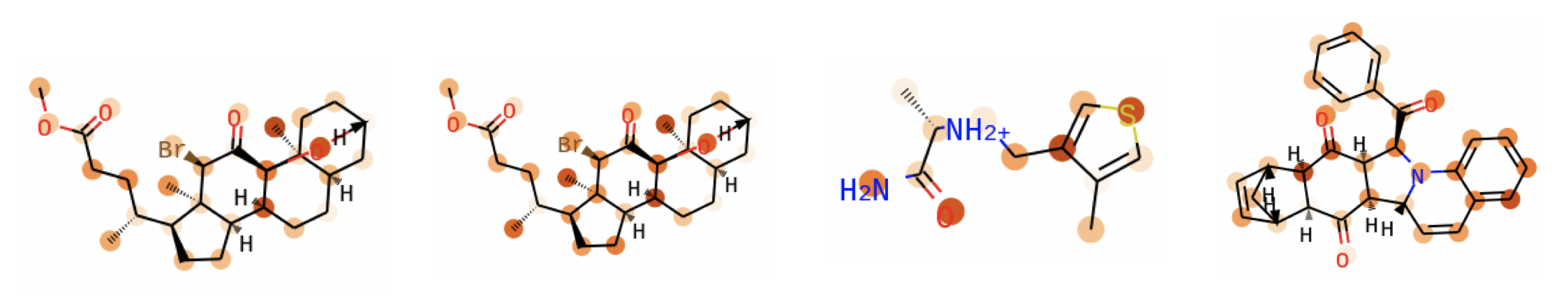}
    \caption{Molecular Group discovery of different chemical compounds.}
    \label{fig:other_highlights}
\end{figure}

Despite only learning the structural representation of the molecule, it learns the active reaction sites as potential functional groups. In the example above, we note that the carbon in aspartate (C) attends less to the $\alpha$-ketoglurate molecule as a whole, whereas the reaction site of [NH3+] attends more strongly to the whole molecule of $\alpha$-ketoglurate. Other visualizations are illustrated as per Figure \ref{fig:other_highlights}.

\section{Multi-task learning for downstream tasks}
\subsection{Generalizing to the graph domain}
After pretraining BERT in an unsupervised manner on ZINC250k, we want to see whether the chemical intuition BERT has picked up is ready to be transferred to other datasets and tasks. In this investigation, we considered the Tox21 dataset, as well as two standard models for inference on it, Graph Convolutional Networks and Graph Attention Networks, comparing different types of featurizations.

\subsubsection{Graph Convolutional Networks}
Letting $R$ be the matrix of features for vertices $r_v$, $A$ be the adjacency matrix of $G$, $D$ be the diagonal matrix where $D_{i,i} = \sum_{j} (A+I)_{i,j}$, and $f(R, A)$ be the function we would like to find that we would like to learn, the $l+1^{\text{st}}$ graph convolutional layer is defined as $$H^{(l+1)}= \sigma\!\left(D^{-\frac{1}{2}} (A+I)D^{-\frac{1}{2}}H^{(l)} W^{(l)} \right)$$ where $H^{(0)}=R$. This model captures the $(l+1)$-hop neighborhood topology of a vertex.

\subsubsection{Graph Attention Networks}
We will describe a single graph attention layer, as it is the sole layer utilized throughout. 

The input to our layer is a set of node features, $h = [h_1,h_2,...,h_N]$,$h_i \in R^F$, where $N$ is the
number of nodes, and $F$ is the number of features in each node. The layer produces a new set of node features after the transformation.

In order to obtain sufficient expressive power to transform the input features into higher-level features, at least one learnable linear transformation is required. Then self-attention is performed on the nodes to compute attention coefficients $e_{ij} = a(Wh_i, Wh_j)$ that indicate the importance of a given node j’s features to node i. In its most general formulation, the model allows every node to attend on every other node, dropping all structural information. The graph structure is injected into the mechanism by performing masked attention---only computing coefficients for nodes based on their neighborhood in the graph. To make coefficients easily comparable across different nodes, they are normalized across all choices of node using the softmax function. 
Once obtained, the normalized attention coefficients are used to compute a linear combination of the features corresponding to them, to serve as the final output features for every node.

\begin{figure}[!htb]
    \centering
    \includegraphics[width=\linewidth]{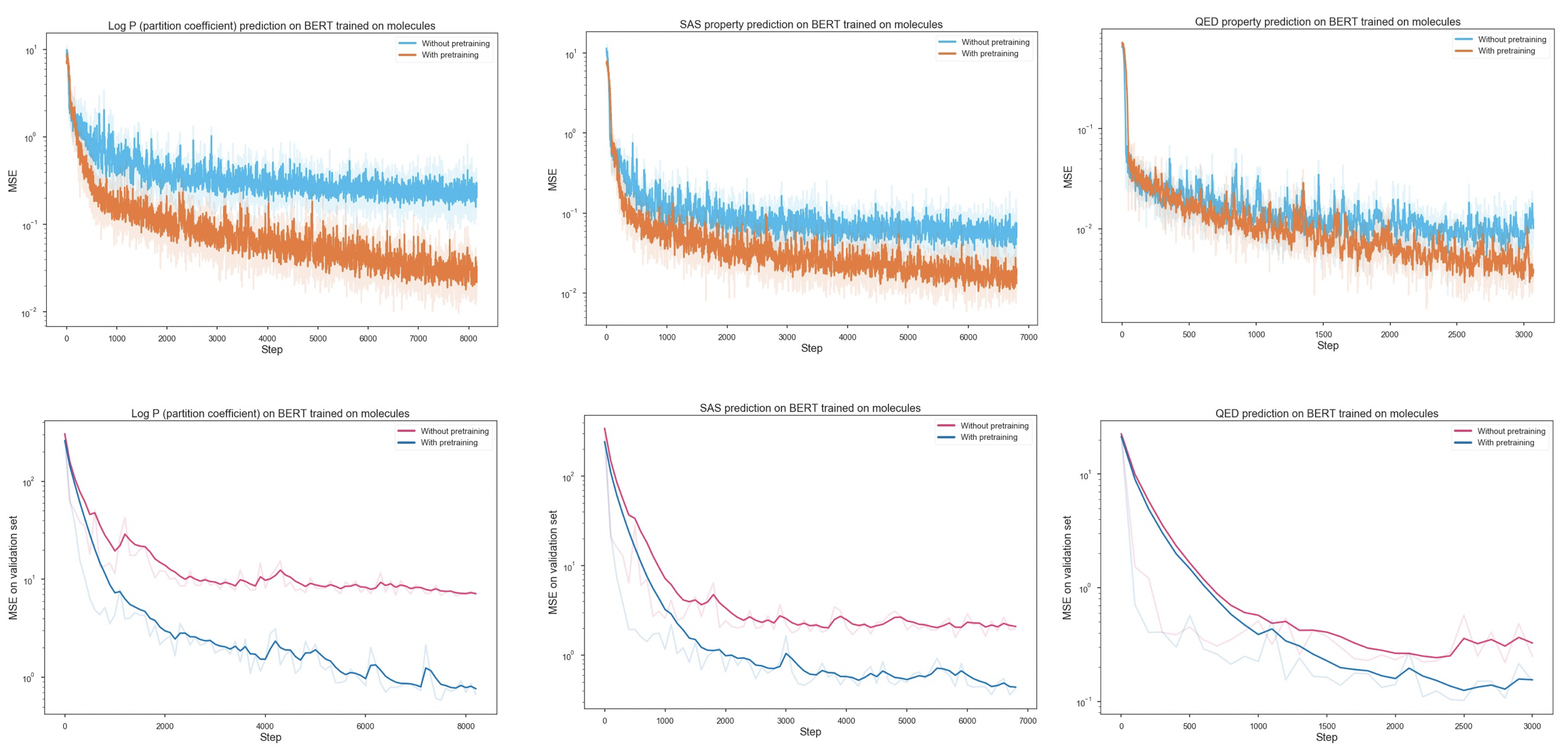}
    \caption{Illustration of BERT used for multi-task learning after pretraining. We test the Zinc250k dataset after pretraining on 3 different task: LogP, the partition coefficient (left), SAS, synthetic accessibility score (middle), QED, the quantitative estimation of drug-likeness (right). Top: training set MSE, bottom: test set MSE.}
    \label{fig:regression}
\end{figure}

\begin{table}[!htb]
  \caption{Comparison of effectiveness of baseline, baseline concatenated with BERT embeddings, and with BERT embeddings only evaluated on GCN and GAT models.}
  \label{table:table-results}
  \centering
  \begin{tabular}{lcc}
    \toprule
     & \multicolumn{2}{c}{Test AUC-ROC} \\
     \cmidrule(r){2-3}
    Model  & GCN & GAT   \\
    \midrule
    Baseline (canonical atom featurization, CAF) & 0.815 & 0.826  \\
    CAF concatenated with BERT embeddings & 0.817 & 0.827 \\
    BERT embeddings only & 0.794 & 0.795 \\
    One-hot encoding of atomic number & 0.790 & 0.788 \\
    Random featurization & 0.678 & 0.673 \\
    \bottomrule
    \end{tabular}
\end{table}
From these results, we found that the embeddings that BERT learns for molecules are not useful in improving upon the state of the art; what it learns seems to be a subset of the information captured by the canonical atom featurizer. One potential detractor from BERT's performance is that there exist molecules in Tox21 with atoms that do not appear in BERT's vocabulary. These are not very many relatively though and should not have a huge impact.
\subsection{Fine-tuning BERT for regression tasks}
Once we pretrained BERT, we decided to fine-tune it with respect to the three values in the ZINC15 dataset. BERT was able to perform this task fairly well, and we found that, intuitively, BERT learning the task with random initial weights did not perform as well as BERT learning the task with pretraining (in both training and validation for all three tasks, as shown in Figure \ref{fig:regression}). 

We also felt that it would be interesting to visualize BERT's attention heads on molecules after it had been fine-tuned to perform a specific chemical inference task. However, we found that the resulting attention differed very little on the whole from task to task, even from pretraining as well. The attention heads displayed seemingly identical behavior for each of the layers across each of the tasks, with the only standout exception being the attention heads for the solubility task in layers 10 and 11. These are the attention heads showcased in Figure \ref{fig:test1}. Some possible causes of this may include the low value range of the QED and SAS tasks (if we unnormalized the regression values, would the gradients be larger, hence causing more obvious change in behavior in the attention heads?) or that BERT's fine-tuning step learns a local optima that is close to the pretraining weights but does not answer anything useful about chemistry. However, the attention heads on the solubility task seemed to behave intuitively, showing that there is a glimmer of hope in this line of investigation. We hope that this is the case, given that a working model like this would be an incredibly helpful tool for chemists.


\section{Future Work and Extensions}
We plan to continue this work using a larger dataset of 96,000,000 molecules gathered from the PubMed dataset\footnote{ftp://ftp.ncbi.nlm.nih.gov/pubchem/Compound/Extras/}, as we've observed that language models like BERT seem to thrive in settings with an abundance of data. Two challenges here are the sheer computational resources required to wield and train on such a massive dataset, and the determination of useful molecules that do not add to the noise in the signal. 

Additionally, we would like to design a loss function for BERT that penalizes distances between embeddings of SMILES strings that represent the same molecule yet are unique. We hope that this will force BERT to learn the inherent graphical meaning of the SMILES strings and hence to start to pick up all of the chemical inference that can be gathered based on connectivity. For this, we will need to augment a smaller dataset, generating multiple unique SMILES strings for the same molecule for each of the molecules in the dataset. At this moment, we are training BERT on such a dataset for future investigation.

Finally, we would like to identify potential flaws in our approach to analyzing chemical property-bearing molecular substructures using fine-tuning. We feel that if implemented correctly, this could be a very useful tool for chemists and could greatly assist drug discovery, reaction prediction, and other tasks.

\section{Discussion and Conclusion}
In this work, we proposed the analysis of BERT and its attention to study functional groups and other property-impacting molecular substructures from a data-driven perspective on datasets of SMILES representations of molecules. We then applied the representations of functional groups and atoms learned by the model to tackle problems of toxicity, solubility, drug-likeness, and synthesis accessibility on smaller datasets using the learned representations as features for graph convolution and attention models on the graph structure of molecules, as well as fine-tuning of BERT. Finally, we proposed the use of attention visualization as a helpful tool for chemistry practitioners and students to quickly identify important substructures in various chemical properties. While we haven't defined the state of the art with any of our results or designed any breakthrough tools, we've discovered several interesting properties of SMILES strings for molecules from an attention model-based standpoint, and our approaches and results may still be useful for others investigating computational chemistry.
\section{Acknowledgements}
We would like to thank Vineet Kosaraju, Michele Catasta, Keiran Thompson, Lillian Zhu, Gaby Li, and Amelia Woodward for their deeply helpful consultation with the project. 
{\small
\bibliographystyle{ieee}
\bibliography{egbib}
}

\end{document}